# Evaluation of Image Noise from Computed Tomography in an Organ-Specific Anthropomorphic Phantom

Lucas McCullum (*0000-0001-9788-7987*)[1,2],
Mojtaba Zarei (*0000-0002-0997-7411*)[3],
Andy Tsai (*0000-0002-0089-6463*)[1,2],
Michael J. Callahan (*0000-0002-7833-4524*)[1,2],
Peng Sun (*0000-0002-9697-4656*)[1,2‡],
Da Zhang (*0000-0001-5092-5421*)[1,2‡*]

[1]Department of Radiology, Boston Children's Hospital, 300 Longwood Avenue, Boston, MA, 02215, USA
[2]Harvard Medical School, 25 Shattuck Street, Boston, MA, 02115, USA
[3]GE HealthCare, Research and Scientific Affairs, Waukesha, WI, 53188, USA

‡ Shared last author
[*]Corresponding Author: Da Zhang <Da.Zhang@childrens.harvard.edu>

**Contributor Roles Taxonomy (CRediT) Attribution Statement:** *Conceptualization*: L.M. and D.Z.; *Data curation*: L.M. and D.Z.; *Formal analysis*: L.M., M.Z., A.T., M.J.C., P.S., and D.Z.; *Funding acquisition*: D.Z.; *Investigation*: L.M. and D.Z.; *Methodology*: L.M., M.Z., and D.Z.; *Project administration*: D.Z.; *Resources*: D.Z.; *Software*: L.M.; *Supervision*: D.Z.; *Validation*: L.M., M.Z., A.T., M.J.C., P.S., and D.Z.; *Visualization*: L.M. and D.Z.; *Writing – original draft*: L.M., M.Z., A.T., M.J.C., P.S., and D.Z.; *Writing - review & editing*: L.M., M.Z., A.T., M.J.C., P.S., and D.Z.;

**IRB Statement:** No human subjects were involved in this study.

**Conflicts of Interest:** None.

**Funding Statement:** None.

**Data Availability Statement:** All relevant anonymized imaging data and source code necessary for figure reproduction and analysis are available at the following FigShare URL: https://doi.org/10.6084/m9.figshare.34913829 and GitHub repository: https://github.com/Lucas-Mc/CT-Noise.

**Acknowledgements:** We would like to thank Kyoto Kagaku for their assistance in creating a customized anthropomorphic phantom.

# Abstract

**Background:** Computed tomography (CT) is used ubiquitously for diagnostic imaging, with novel iterative reconstructions (IR) such as hybrid iterative reconstruction with partial model-based component (ASiR-V) and deep learning image reconstruction (DLIR) increasingly utilized. In conjunction, CT vendors such as GE HealthCare have developed a noise index (NI) to reduce radiation dose during protocol development. However, it is unclear how these reconstruction algorithms reduce noise as NI changes, particularly across multiple organs.

**Purpose:** The goal of this study was to utilize an anthropomorphic phantom to: (1) evaluate image noise against the prescribed NI across different organs, (2) determine if DLIR reduces image noise compared to ASiR-V, and (3) use this relationship to determine the maximum volume CT dose index ($CTDI_{vol}$) reduction at equivalent noise levels.

**Methods:** Repeat images were acquired on a GE HealthCare Revolution Apex CT scanner across NI values in an anthropomorphic adult-sized phantom from Kyoto Kagaku with ASiR-V and DLIR. Organs-of-interest were contoured using TotalSegmentator in 3D Slicer. The correlation between the $CTDI_{vol}$ and NI was determined. The standard deviation of the difference in the repeat images was used as a noise metric (NM), correlated with NI, used to determine the NI at equivalent NM between ASiR-V and DLIR, then applied to determine the dose reduction and difference in NM.

**Results:** The relationship between NI and the resulting $CTDI_{vol}$ was found to be proportional to a power law of exponent -2.23. The NM was dependent on NI using a power law with exponent under 1 with smaller exponents for DLIR. The maximum potential absolute dose reduction from using DLIR was found in the liver at 1.3 [1.0 to 1.8, 95% confidence interval] mGy (approximately 22% of the corresponding ASiR-V $CTDI_{vol}$) with an NI of 7.9 [6.4 to 9.3, 95% confidence interval]. The spinal cord and gallbladder were least impacted by DLIR. Across all organs, the relative dose reduction increased, the NM-to-NI ratio decreased, and the noise reduction increased across NI.

**Conclusions:** The relationship between image noise and the prescribed NI differed between ASiR-V and DLIR and varied by organ, with model-based dose reductions of up to 1.3 mGy (approximately 22%) in the liver. Because no task-based or observer assessment was performed, these estimates are noise-matched projections rather than validated protocol recommendations and warrant further investigation to optimize CT protocols with both dose and diagnostic purpose in mind.

## Graphical Abstract

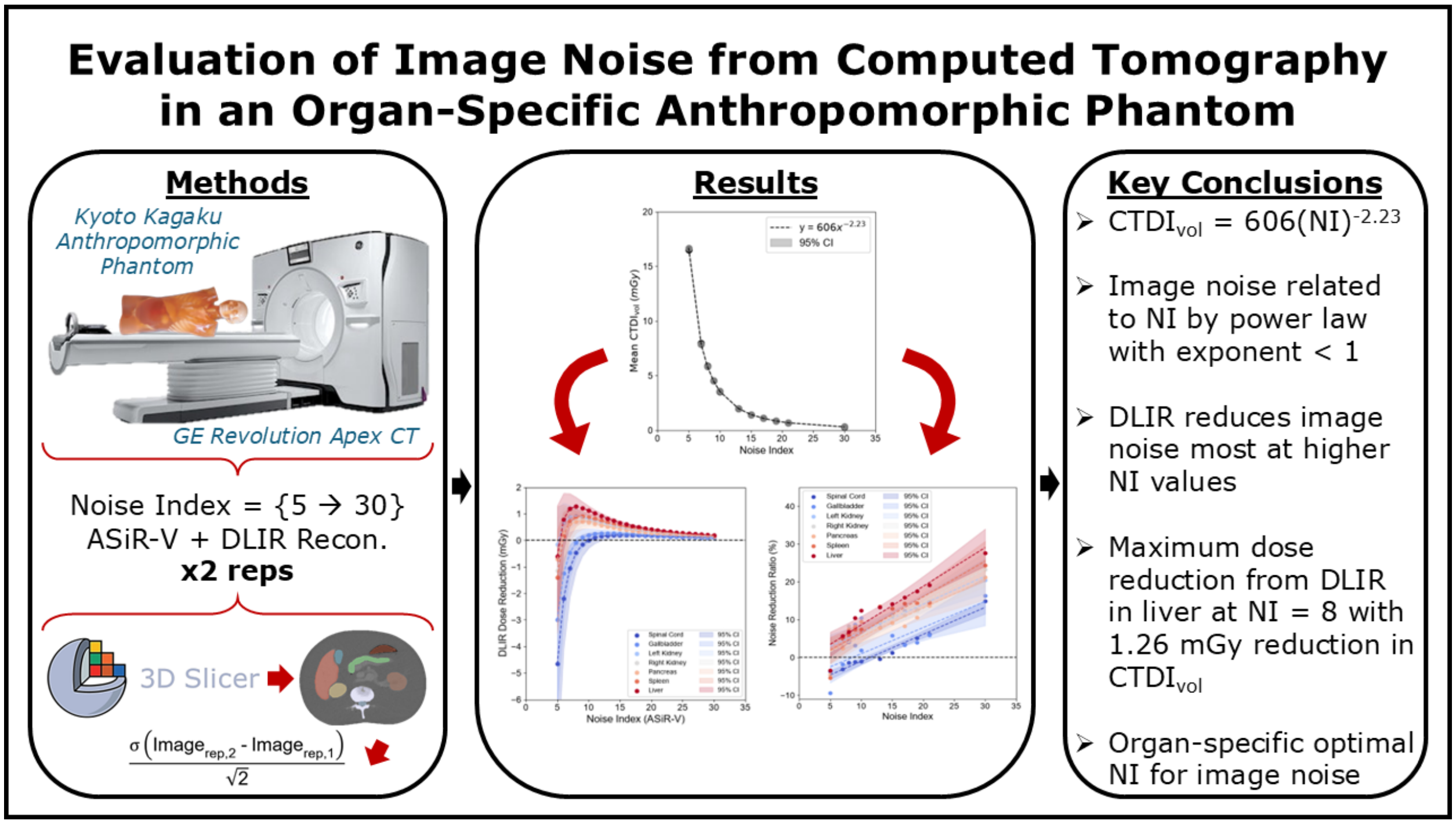

# 1. Introduction

Computed tomography (CT) has become a mainstay in current clinical practice due to its fast acquisition times, particularly in the pediatric setting. These fast acquisition times are also driven by CT's use of ionizing radiation which should be minimized while achieving the required image quality, especially in the pediatric setting. In modern CT systems, automatic exposure control (AEC) is used to attempt to maintain either a constant image noise or surrogate of image quality across the entire scan range in which the patient thickness and internal total attenuation may vary dramatically. Thus, utilizing the full functionality of AEC technology can considerably reduce the total dose to the patient. GE HealthCare (Waukesha, WI, USA) CT scanners have two primary ways of performing AEC during scanning, both of which require a single scout image to determine the required tube output adjustments. The first is "AutomA" which performs tube current modulation along the longitudinal direction while the second is "SmartmA" which performs tube current modulation along both the longitudinal and angular directions[1]. The level of tube output is determined by the prescribed "Noise Index (NI)" which is used as a surrogate for total image noise. During both AutomA and SmartmA, the system attempts to maintain a constant level of image noise across the entire scan range by performing the respective tube current modulations. This value can be prescribed by the scanner operator where lower values indicate lower overall image noise. When evaluating the tube current from the digital imaging and communications in medicine (DICOM) tags, since images are saved per-slice and SmartmA applies both longitudinal and angular tube current modulation, the tube current saved for each slice is an effective value averaged across the relevant projection angles.

Another factor that influences CT image noise is the specific reconstruction method used. Traditionally, filtered back projection[2] (FBP), and later iterative reconstructions[3] such as adaptive statistical iterative reconstruction[4] (ASiR), were the most common reconstruction techniques used for CT scans. Following this, model-based iterative reconstruction (marketed as Veo for GE HealthCare scanners) reduced noise even further but at a high computational burden[5]. To reduce the computational burden, GE HealthCare released ASiR-V which combines the speed of traditional ASiR with the reduced noise capabilities of Veo to further enhance online reconstructions[6,7]. However, these reconstruction methods have recently been challenged by deep learning image reconstruction[8] (DLIR / TrueFidelity, GE HealthCare) which promises to reduce image noise at the same radiation dose. Both ASiR-V and DLIR can be used at different strength levels. For ASiR-V, the strength level adjusts the percentage blend of conventional filtered back projection reconstruction compared to the novel iterative methods. For DLIR, the low, medium, and high strength levels correspond to different degrees of noise reduction produced by the trained deep-learning reconstruction algorithm and are not defined as linear blending ratios with conventional reconstruction. Since DLIR is a recent development with many clinical centers only recently adopting this technology, its relationship to the prescribed NI is largely unknown especially on a per-organ basis.

A common method for evaluating CT dosimetry is the volume CT dose index ($CTDI_{vol}$) which approximates the equivalent dose of a helical CT scan to that of an axial CT scan for easier dose comparisons. This value must be displayed on the CT scanner for each acquisition and is derived from dosimetry measurements in a cylindrical phantom[9]. One limitation of the cylindrical CT phantom is that it does not accurately represent the potential dosimetry and resulting image quality effects of specific organs within the body which creates a gap for clinical translation of CT protocol development. Recently, Kyoto Kagaku (Kyoto, Japan) has developed anthropomorphic phantoms which can be imaged using CT scanners that provides realistic models of organs and their associated expected CT numbers[10].

While scanner and organ agnostic studies have been explored to study the optimal tradeoffs of image noise and dose[11], limited work has employed this methodology to study the behavior specifically on the GE HealthCare CT scanners using an anthropomorphic phantom which can realistically simulate noise behavior across organs. Therefore, the goal of this study is to utilize the novel realistic anthropomorphic phantom from Kyoto Kagaku to: (1) evaluate image noise performance against prescribed NI across different anatomical regions of interest, (2) determine if the use of DLIR is beneficial at reducing image noise compared to ASiR-V reconstruction across different prescribed NI values and anatomical regions of interest, and (3) use this relationship to determine the maximum potential dose reductions at equivalent noise levels on a per-organ basis.

# 2. Methods and Materials

## 2.1. CT Acquisition Parameters

All data were acquired using a GE HealthCare Revolution Apex CT (software version: 24MW10.61). The scan range was set from the mid-pelvis to the mid-lung to assess a wide range of abdominal anatomy. The set of NI values of 5, 7, 8, 9, 10, 13, 15, 17, 19, 21, and 30 were chosen for this study. The focal spot size was 1.5 / 1.8 mm for NI = 5, 1.2 / 1.6 mm for NI = 7, and 0.7 / 1.0 mm for the remaining acquisitions. The average tube current across all slices was reported due to the use of SmartmA. A summary of the final acquisition parameters can be seen in **Table 1** while a visualization of the varying tube current can be seen in **Supplementary Figure S1**. Two different reconstructions were used and were generated from the same raw data acquisition: (1) ASiR-V with 50% strength and (2) DLIR with medium strength, both with the standard convolution kernel.

**Table 1.** The CT acquisition parameters for this study. Apart from the focal spot size, which differed at NI = 5 and NI = 7 (see Section 2.1), the only setting that was changed across acquisitions was the prescribed NI.

| | Setting | | | | | | | | | | |
|---|---|---|---|---|---|---|---|---|---|---|---|
| Tube Voltage (kVp) | 120 | | | | | | | | | | |
| NI | 5 | 7 | 8 | 9 | 10 | 13 | 15 | 17 | 19 | 21 | 30 |
| Minimum Tube Current (mA) | 546 | 264 | 197 | 153 | 121 | 69 | 50 | 39 | 30 | 24 | 11 |
| Average Tube Current (mA) | 690 | 328 | 243 | 187 | 147 | 82 | 59 | 45 | 35 | 28 | 12 |
| Maximum Tube Current (mA) | 837 | 397 | 291 | 223 | 173 | 94 | 67 | 51 | 39 | 31 | 13 |
| Exposure Time (ms) | 500 | | | | | | | | | | |

| | |
|---|---|
| Filter Type | Large Filter |
| Detector Type | 128 x 0.625 mm |
| Rotation Time (s) | 0.5 |
| Pitch | 1.375 |
| Acquired Voxel Size ($mm^3$) | 0.68 x 0.68 x 5 |
| Acquired Scan Length (mm) | 356.5 |
| Reconstructed Matrix Size | 512 x 512 x 55 |
| Reconstructed Scan Length (mm) | 275 |
| Reconstruction Diameter (mm) | 350 |
| Reconstructed Slice Gap (mm) | 0 |
| Total Scan Time (s) | 1.62 |

### 2.2. Phantom Assessment

A custom-built, adult-sized, anthropomorphic phantom from Kyoto Kagaku based on the PBU-60 model was used for all image acquisitions. The spinal cord, gallbladder, left kidney, right kidney, pancreas, spleen, and liver were contoured using a single repetition of the NI = 5 images using the "total" segmentation task at normal speed in TotalSegmentator[12,13] version 2.14.0 using 3D Slicer[14] version 5.12.3 (https://www.slicer.org). The NI = 5 images were chosen since they had the lowest level of image noise. All segmentations were reviewed by an experienced medical professional and manually refined with minor adjustments as needed to ensure complete organ coverage, particularly when the automatic segmentation did not fully capture the organ. The remaining segmentations were not used either due to only partially being inside the imaging volume (i.e., lung) or lack of confidence with accurate CT numbers (i.e., stomach). A visualization of the resulting chosen segmentations on the acquired NI = 5 CT image can be seen in **Figure 1** and a table of their volume, approximate depth, and CT number for both the ASiR-V and DLIR reconstruction is given in **Supplementary Table S1**.

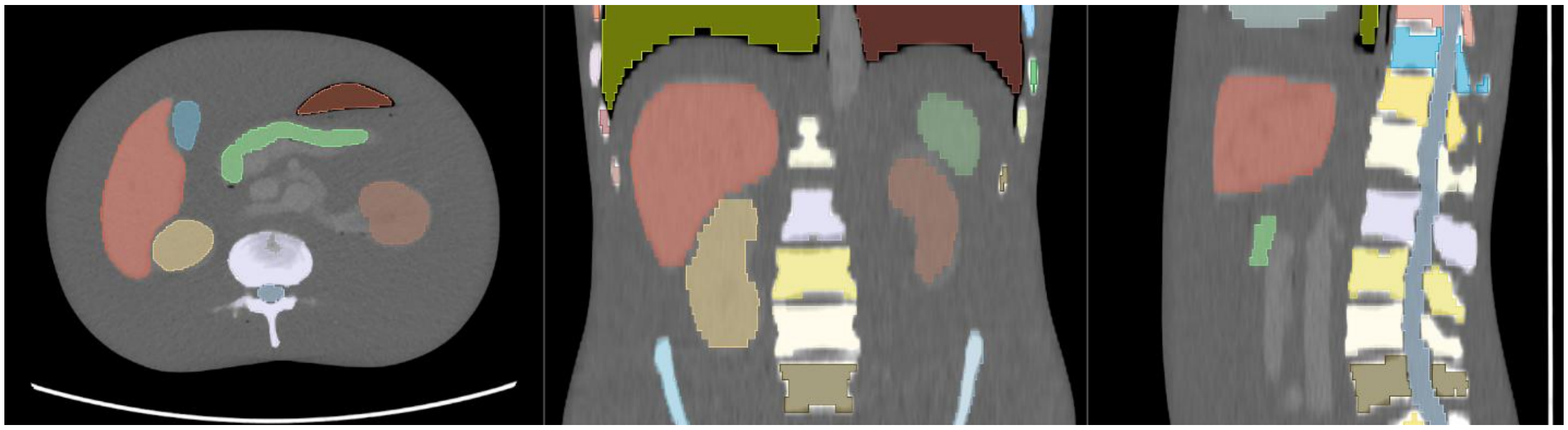

**Figure 1.** An overview of the resulting image segmentations from TotalSegmentator in the axial (left), coronal (middle), and sagittal (right) planes overlayed on the acquired CT with the lowest prescribed NI of 5. Also shown here are the segmented ribs and vertebrae which were not analyzed in this study.

Since all images were acquired in the same frame of reference and without moving the phantom, no image registration or resampling had to be done to apply the segmentation to the other acquisitions.

### 2.3. Statistical Analysis

All computational analysis was completed using Python 3.12.10. To assess the repeatability of the consecutive acquisitions, the coefficient of variation was used for the tube current values and averaged across each slice and compared across prescribed NI values. Several metrics were computed in this study to evaluate the effects of image noise across anatomical regions. The first was the noise metric (NM) which was defined as the standard deviation of the subtraction image across the two repetitions divided by the square root of 2 to account for noise propagation[15]. Edge features were not excluded from the noise measurements to preserve their effects on acquired and reconstruction produced noise. Note that the NM may contain contributions from organ boundaries, partial-volume effects, and reconstruction-dependent edge variability in addition to stochastic image noise.

$$\text{Noise Metric (NM)} = \frac{\sigma\left(\text{Image}_{rep,2} - \text{Image}_{rep,1}\right)}{\sqrt{2}}$$

The second was the NM-to-NI ratio which was defined as the computed NM divided by the prescribed NI. The third was the noise reduction ratio which was defined as the percent difference between the NM for the DLIR image against the ASiR-V image.

$$\text{Noise Reduction Ratio} = 100\% * \frac{NM_{ASiR\text{-}V} - NM_{DLIR}}{NM_{ASiR\text{-}V}}$$

The relationship between the noise reduction ratio and the prescribed NI was modeled using a linear fit while all other relationships were modeled using a non-linear least squares on the log-transformed data with no weighting on a power law with variable amplitude and exponent with the fit constrained to pass through the origin (no additive offset term was estimated). For the linear fitting method, the coefficient of determination[16] ($r^2$) was calculated to assess the quality of the fitting. After computing the NM for each reconstruction, its value for the DLIR reconstruction was plotted against the ASiR-V reconstruction and fit using a power law of the following form. A linear fit was also tried initially, but this did not fully capture the non-linear trends seen.

$$NM_{DLIR} = a_{NM}(NM_{ASiR\text{-}V})^{b_{NM}}$$

With this relationship derived, it can be combined with the previous NM versus NI ASiR-V fits and then used to determine the equivalent NI for the DLIR reconstruction that would result in the same NM as the ASiR-V reconstruction. After the relationship between the NI of the two reconstructions was established, the respective $CTDI_{vol}$ was computed as a way to determine the potential reductions in scanner x-ray tube radiation output using the previously fit relationship for the ASiR-V reconstruction. Both of these derivations can be seen in the **Supplementary Materials**. From this relationship, the respective model-based dose reduction from using DLIR for each organ could be determined. For each relationship, 95% confidence intervals (CI) were generated and were carried forward into subsequent calculations, if necessary. The 95% confidence intervals were computed by 10,000 simulated power law curves with the amplitudes and exponents sampled from a multivariate normal distribution with the respective covariance matrix.

# 3. Results

A demonstration of the acquired images is shown in **Figure 2**. As the NI increases, the apparent noise in the image increases, particularly in the central region of the image. This is best appreciated in the bottom row of the figure which shows the difference between consecutive repetitions at the same NI value. Interestingly, the degree of noise along the edges in the image become more pronounced as the NI increases, except for when NI = 5 which shows strong differences at the edges but notably reduced differences in the liver compared to the higher NI value.

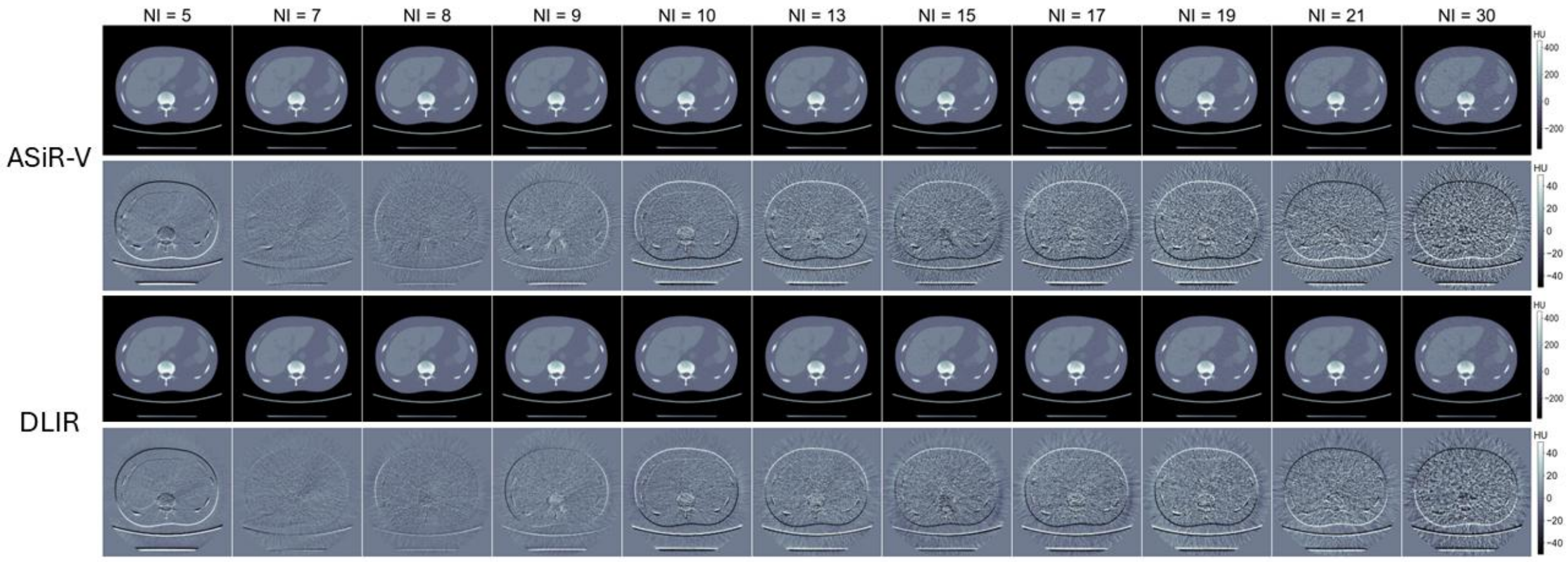

**Figure 2.** A visualization of one repetition of each prescribed NI value (top) and the absolute difference between the two repetitions (bottom) for both ASiR-V (top two rows) and DLIR (bottom two rows).

A plot demonstrating the change in mean $CTDI_{vol}$ as a function of the prescribed NI with a power law line-of-best-fit is shown in **Figure 3**. The relationship between the prescribed NI value and the resulting $CTDI_{vol}$ can be approximated using this combination of phantom and acquisition parameters by a magnitude of 606 (95% CI = [594, 617]) and power of -2.23 (95% CI = [-2.24, -2.22]). The repeatability in the average tube current was calculated to be less than 0.4% in terms of the coefficient of variation for all prescribed NI values.

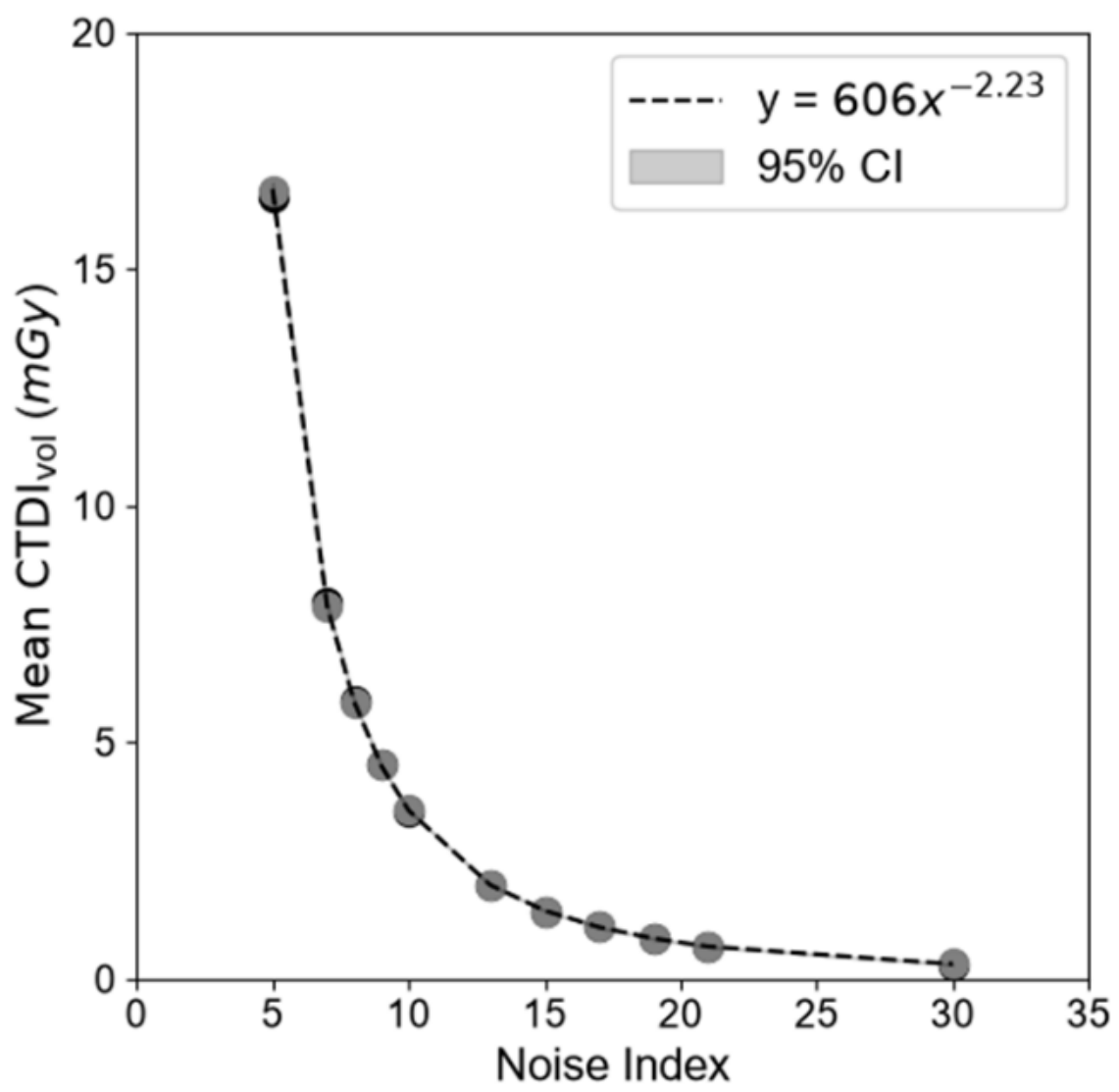

**Figure 3.** The relationship between the mean $CTDI_{vol}$ and the prescribed NI (black and grey circles) with overlayed power law curve fitting (dashed line) and respective 95% CI (shaded region) using both repetitions. Both repetitions are displayed here and represented as a black and grey circle but are difficult to differentiate due to their high repeatability.

When comparing our calculated NM to the prescribed NI, a consistent positive association was found between the two values across each anatomical region as shown in **Figure 4**.

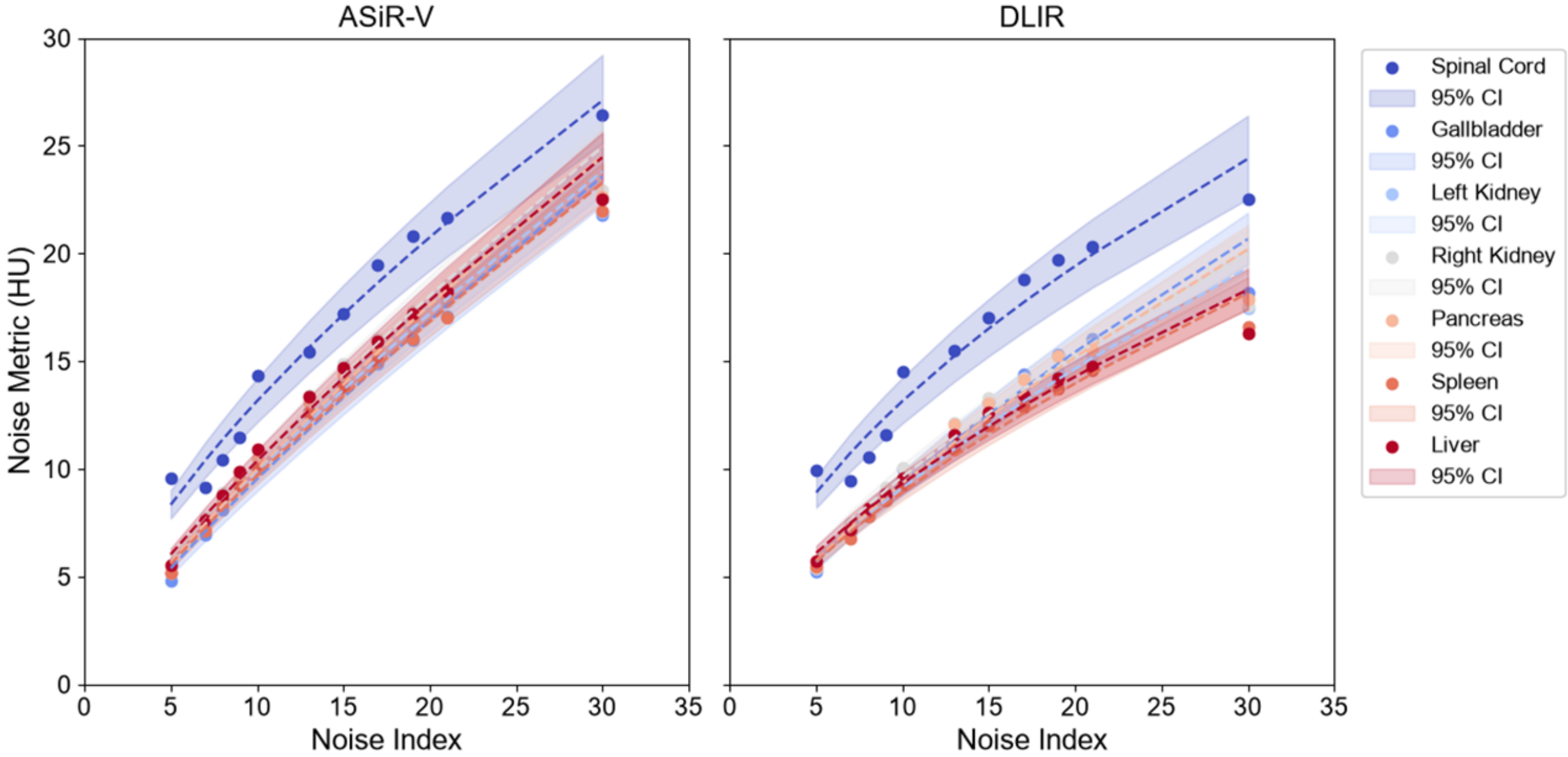

**Figure 4.** The relationship between the computed NM and the prescribed NI for each anatomical region of interest using both ASiR-V (left) and DLIR (right) with overlaying power law curve fitting (dashed lines) and 95% CI (shaded region).

The resulting quantitative values for the power law magnitude and exponent for each line-of-best-fit are shown in **Supplementary Table S2** and separated by reconstruction technique and anatomical region. The computed NM was generally lower across all prescribed NI values for the DLIR reconstruction compared to the ASiR-V reconstruction. This relationship can be seen further in **Figure 5** with the respective fitting results in **Supplementary Table S3**.

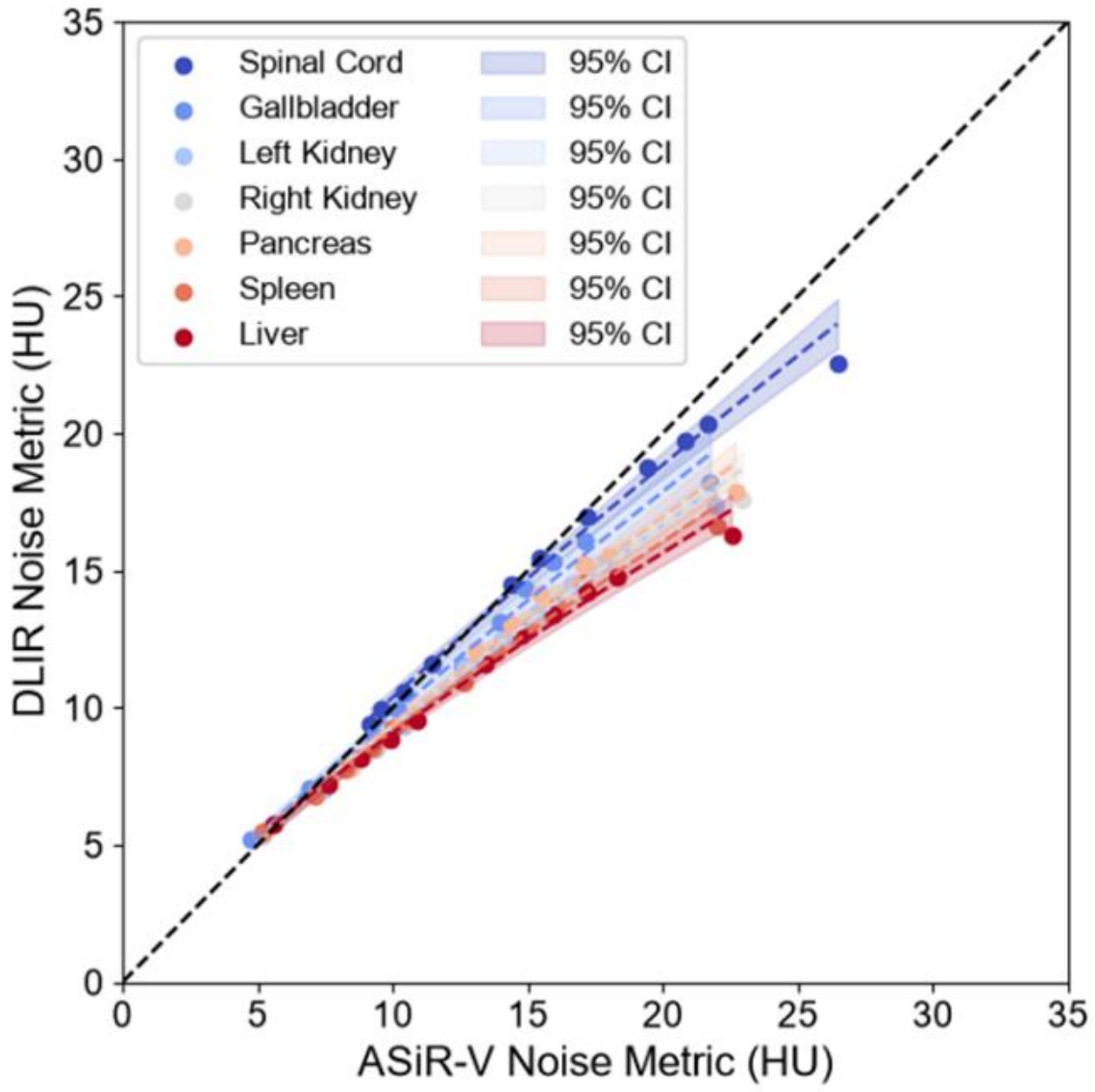

**Figure 5.** The relationship between the computed NM for the DLIR reconstruction compared to the ASiR-V reconstruction at equivalent prescribed NI values with overlying power law curve fitting (dashed lines) and 95% CI (shaded region).

As shown in **Figure 5**, the computed NM at equivalent prescribed NI values has the greatest reduction in the liver and the least reduction in the spinal cord with greater reduction seen at higher computed NM values. The equivalent prescribed NI value for the DLIR reconstruction that would result in the same computed NM as the ASiR-V reconstruction is given in **Figure 6** using the derived fitting relationships from **Figure 4** and **Figure 5**.

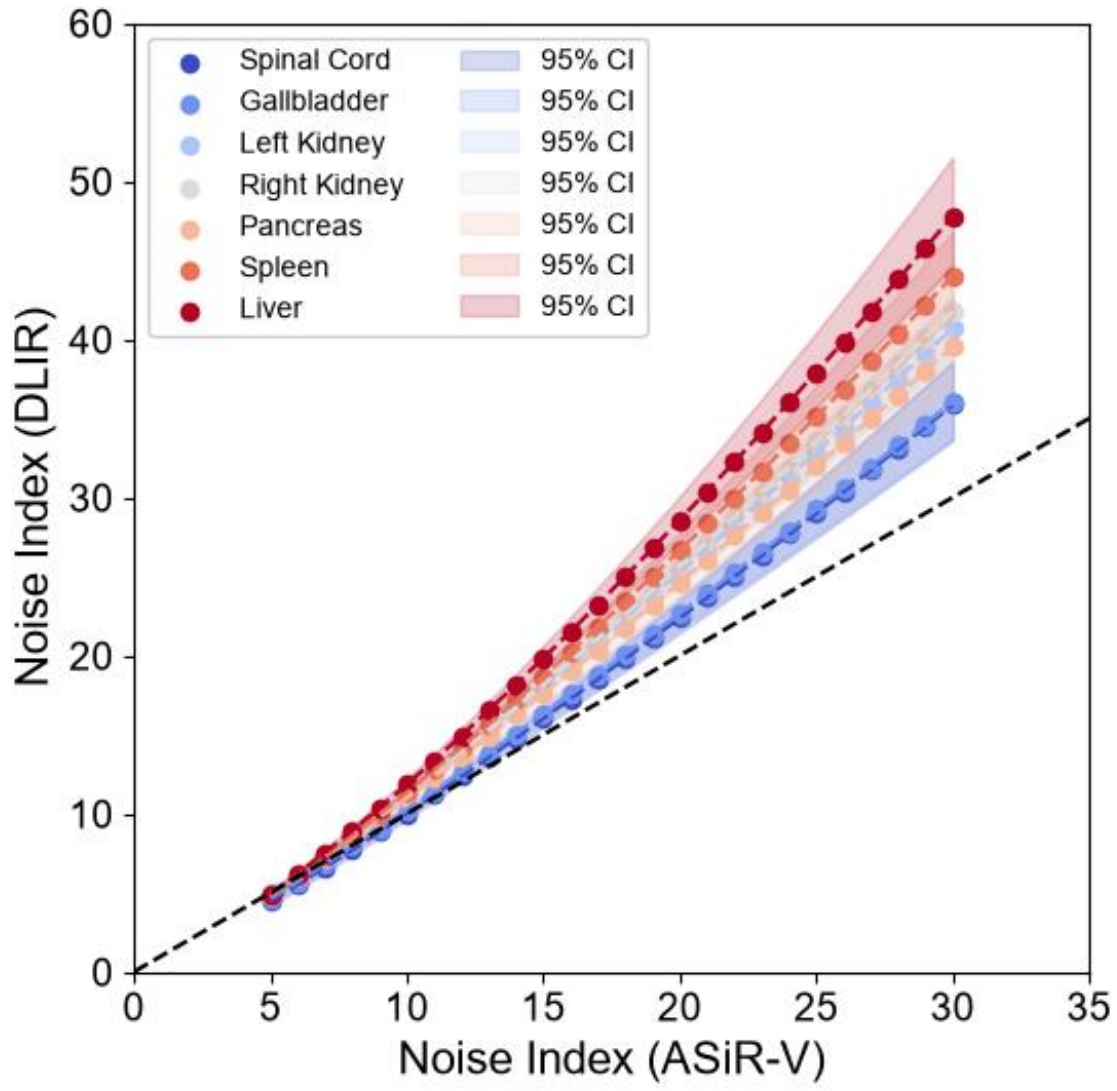

**Figure 6.** The relationship between the equivalent prescribed NI value for the DLIR reconstruction that would result in the same computed NM value as the ASiR-V reconstruction with overlying power law curve fitting (dashed lines) and 95% CI (shaded region).

It can be seen from this figure that the benefit of using the DLIR reconstruction increases as the prescribed NI for the ASiR-V reconstruction increases. However, at low prescribed NI values, there may be a slight detrimental effect when using the DLIR reconstruction. The benefit is seen the most in the liver while the detrimental effect is seen the most in the spinal cord. The resulting impacts on dose as measured using the $CTDI_{vol}$ can be seen in **Figure 7**.

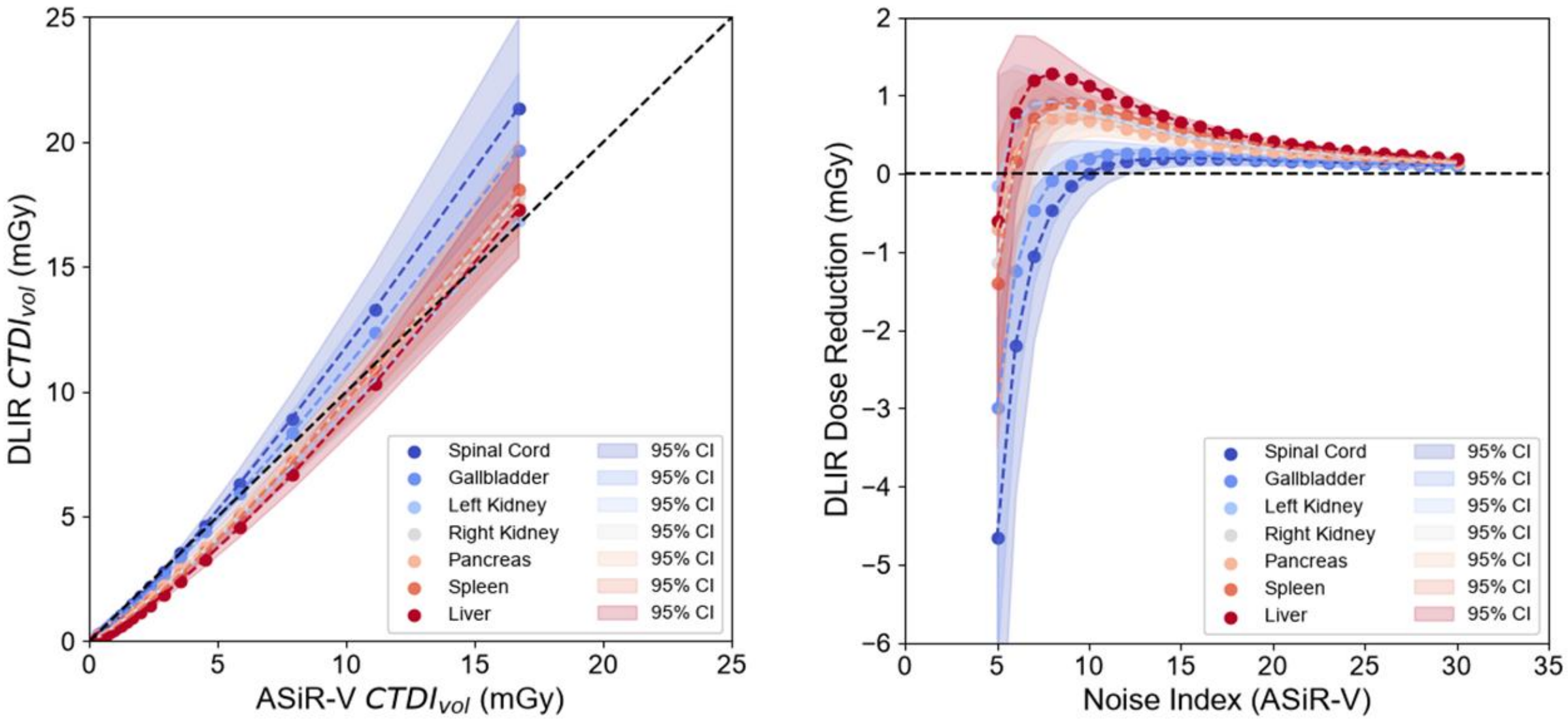

**Figure 7.** The relationship between the $CTDI_{vol}$ for the DLIR reconstruction to achieve the same computed NM as the ASiR-V reconstruction (left) and the resulting dose reduction by switching to DLIR reconstruction (right) with overlying curve fitting (dashed lines) and 95% CI (shaded region).

The dose reduction possible when using DLIR reconstruction to achieve the equivalent NM as the ASiR-V reconstruction is maximized in the liver with a reduction of 1.3 mGy and minimized in the spinal cord with a reduction of 0.2 mGy as shown quantitatively in **Table 2**.

**Table 2.** Relationship between the dose reduction (mGy) from using the DLIR reconstruction to achieve the same computed NM as the ASiR-V reconstruction along with the maximum dose reduction and its respective prescribed NI value.

| | **NI for Maximum Dose Reduction (95% CI)** | **Maximum Dose Reduction (95% CI) (mGy)** |
|---|---|---|
| Spinal Cord | 15.0 (11.6, 18.3) | 0.2 (0.1, 0.3) |
| Gallbladder | 12.7 (9.0, 15.7) | 0.3 (0.2, 0.4) |
| Left Kidney | 7.7 (6.0, 9.3) | 0.9 (0.7, 1.4) |
| Right Kidney | 8.7 (6.5, 10.6) | 0.8 (0.6, 1.3) |
| Pancreas | 8.5 (5.6, 10.8) | 0.7 (0.5, 1.4) |
| Spleen | 8.7 (7.2, 10.1) | 0.9 (0.7, 1.2) |
| Liver | 7.9 (6.4, 9.3) | 1.3 (1.0, 1.8) |

For all organs except the gallbladder and spinal cord, the prescribed NI to achieve a maximum dose reduction by switching to the DLIR reconstruction from the ASiR-V reconstruction was 7.9 to 8.7 while the optimal NI for the gallbladder is 12.7 and 15.0 for the spinal cord. The relationship between the noise reduction ratio and the prescribed NI is shown in **Figure 8**. All relationships were positive, meaning that increasing the prescribed NI increases the noise reduction capabilities of the DLIR reconstruction compared to the ASiR-V reconstruction. This effect was the most enhanced in the liver while seen at higher prescribed NI values in the spinal cord and gallbladder.

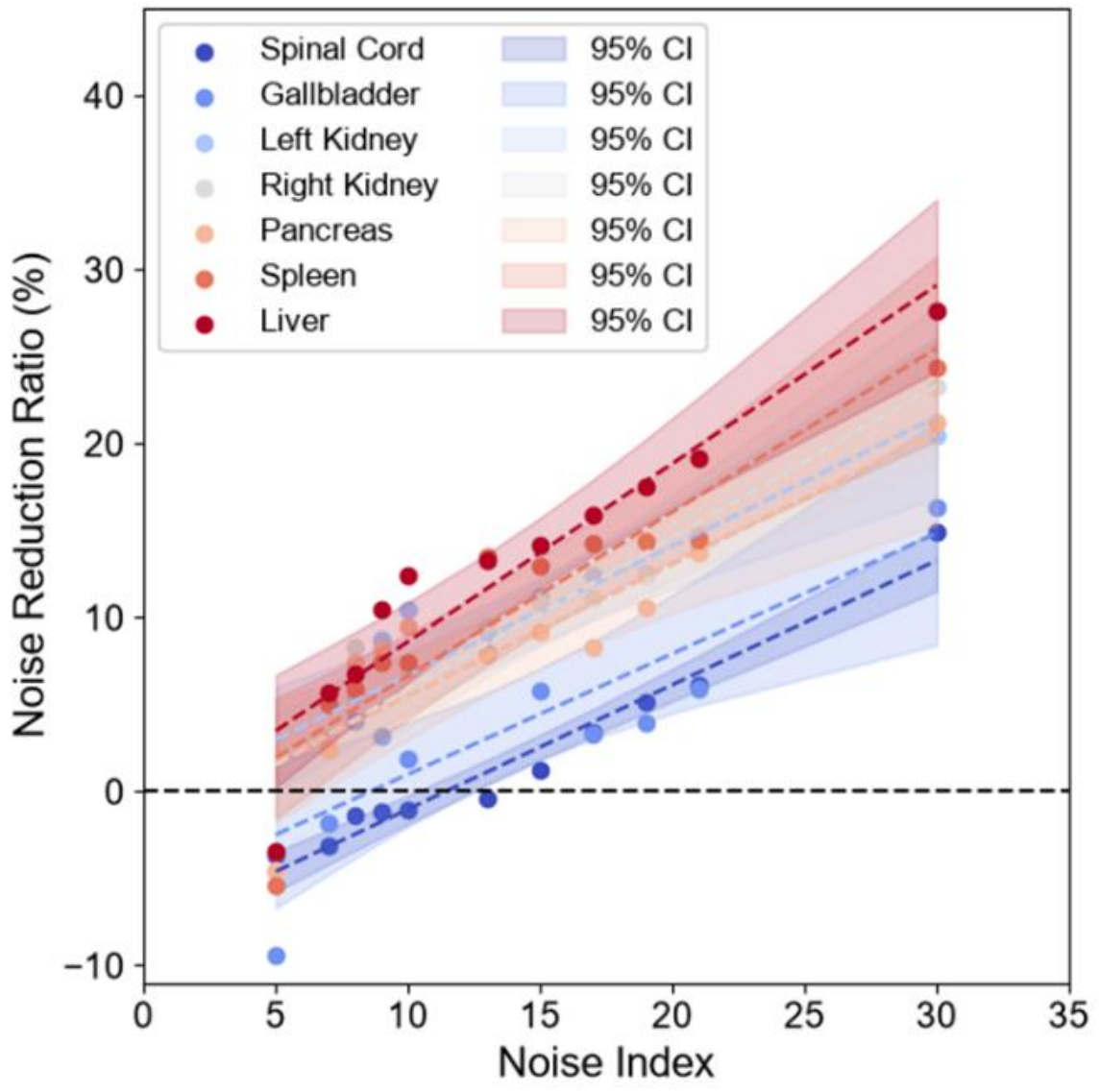

**Figure 8.** The relationship between the noise reduction ratio and the prescribed NI for each anatomical region with accompanying linear lines-of-best-fit (dashed lines) and 95% CI (shaded region).

The detailed numerical results for **Figure 8** are shown in **Table 3**. As shown, the DLIR reconstruction provided a noise reduction compared to the ASiR-V reconstruction at all tested prescribed NI values in the left kidney, right kidney, pancreas, spleen, and liver while the cutoff was greater than 8 for the gallbladder and greater than 11 for the spinal cord. The steepest slope was seen in the liver at 1.02 while the shallowest was seen in the gallbladder at 0.69 indicating their susceptibility to noise reduction as the prescribed NI increases. A general trend was seen where the organs with the higher CT numbers exhibited a lower cutoff for the prescribed NI to achieve a positive noise reduction ratio which may also be affected by the organ's depth, volume, and surrounding structures.

**Table 3.** Relationship between the noise reduction ratio (%) for the DLIR reconstruction compared to the ASiR-V reconstruction across the prescribed NI values. Also shown are the cutoff prescribed NI values which result in the DLIR reconstruction reducing the image noise

compared to the ASiR-V reconstruction. The equation a*NI + b and respective 95% CI for the cutoff NI are also shown. *These values lie below the tested minimum NI domain of 5 and should therefore be used with caution as they represent extrapolations.

| | a (95% CI) | b (95% CI) | $r^2$ | Cutoff NI (95% CI) |
|---|---|---|---|---|
| Spinal Cord | 0.72 (0.61, 0.82) | -8.18 (-9.76, -6.61) | 0.97 | >11.4 (10.8, 12.0) |
| Gallbladder | 0.69 (0.33, 1.06) | -5.96 (-11.72, -0.19) | 0.67 | >8.6 (0.6*, 11.0) |
| Left Kidney | 0.74 (0.48, 1.01) | -0.73 (-4.88, 3.42) | 0.82 | >1.0* (-7.2*, 4.8) |
| Right Kidney | 0.87 (0.61, 1.13) | -2.84 (-6.95, 1.26) | 0.86 | >3.3* (-2.1*, 6.1) |
| Pancreas | 0.76 (0.45, 1.06) | -2.07 (-6.83, 2.70) | 0.78 | >2.7* (-6.0, 6.4) |
| Spleen | 0.94 (0.64, 1.24) | -2.79 (-7.52, 1.95) | 0.85 | >3.0* (-3.0*, 6.0) |
| Liver | 1.02 (0.74, 1.31) | -1.65 (-6.08, 2.78) | 0.88 | >1.6* (-3.8*, 4.6*) |

When comparing the calculated NM-to-NI ratio to the prescribed NI, a consistent negative association was seen across each anatomical region as shown in **Figure 9** with detailed numerical results in **Supplementary Table S2** where the exponent is subtracted by 1 since the NM fit is divided by NI.

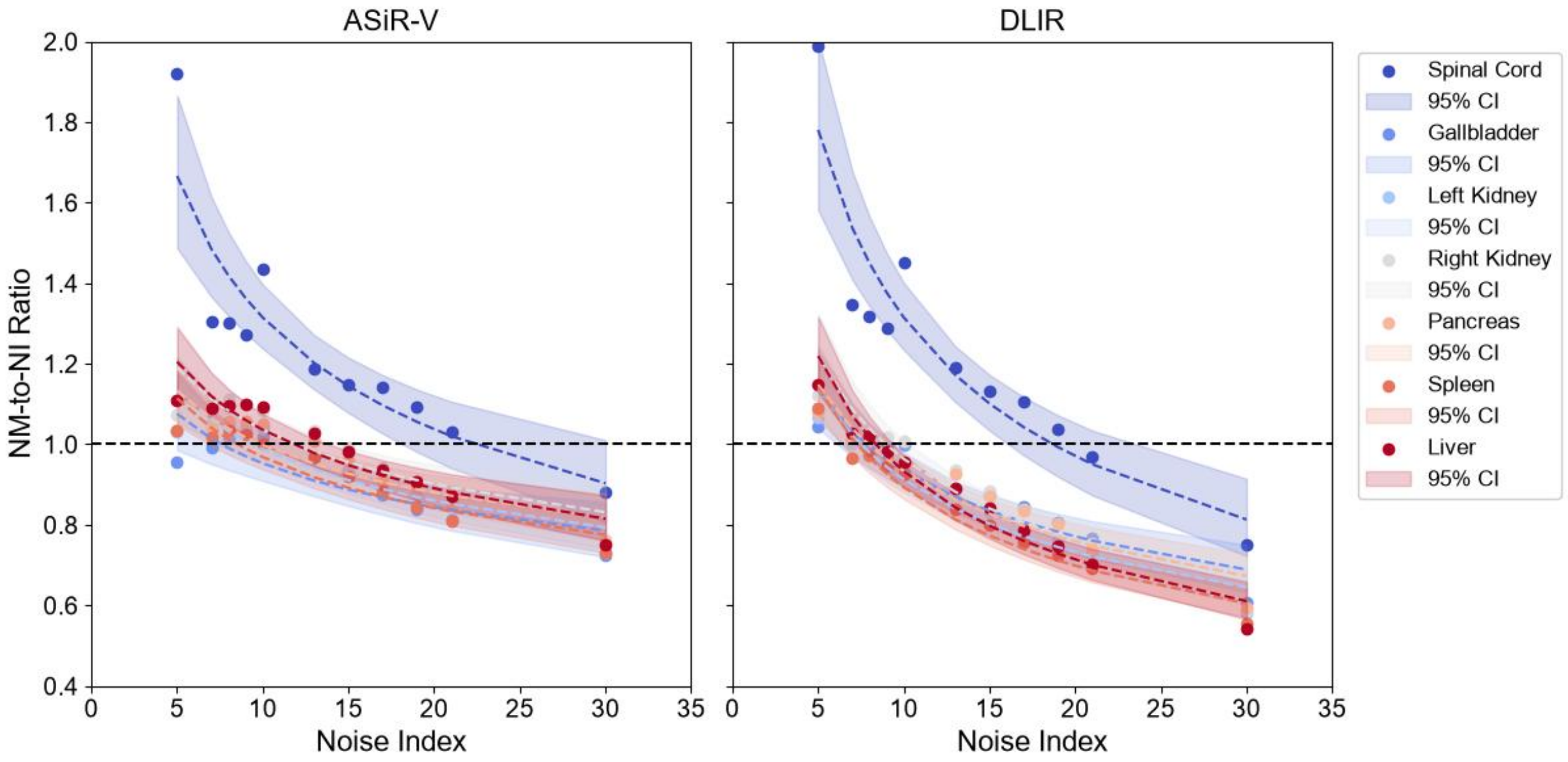

**Figure 9.** The relationship between the computed NM-to-NI ratio and the prescribed NI for each anatomical region of interest using both ASiR-V (left) and DLIR (right) with accompanying power lines-of-best-fit (dashed lines) and 95% CI (shaded region).

A consistent negative trend between the NM-to-NI ratio and the prescribed NI value can be seen for both the ASiR-V and DLIR reconstructions across all tested organs. The spinal cord had the highest average NM-to-NI ratio along with the steepest decrease in NM-to-NI ratio as the prescribed NI increased. Interestingly, for NI values under 10, consistent NM-to-NI ratios near 1 were seen for all tested organs except the spinal cord in the ASiR-V reconstructions while a monotonic decrease was seen for DLIR. When comparing the NM-to-NI ratio from the DLIR reconstruction to the ASiR-V reconstruction at the same prescribed NI value, at the lower ASiR-V NM-to-NI ratio values, the DLIR reconstruction had a lower NM-to-NI ratio than ASiR-V while becoming higher as the ASiR-V NM-to-NI ratio increased. This reversal was seen around a NM-to-NI ratio of approximately 1.2. A visualization can be seen in **Figure 10** with resulting fitting results in **Supplementary Table S4**.

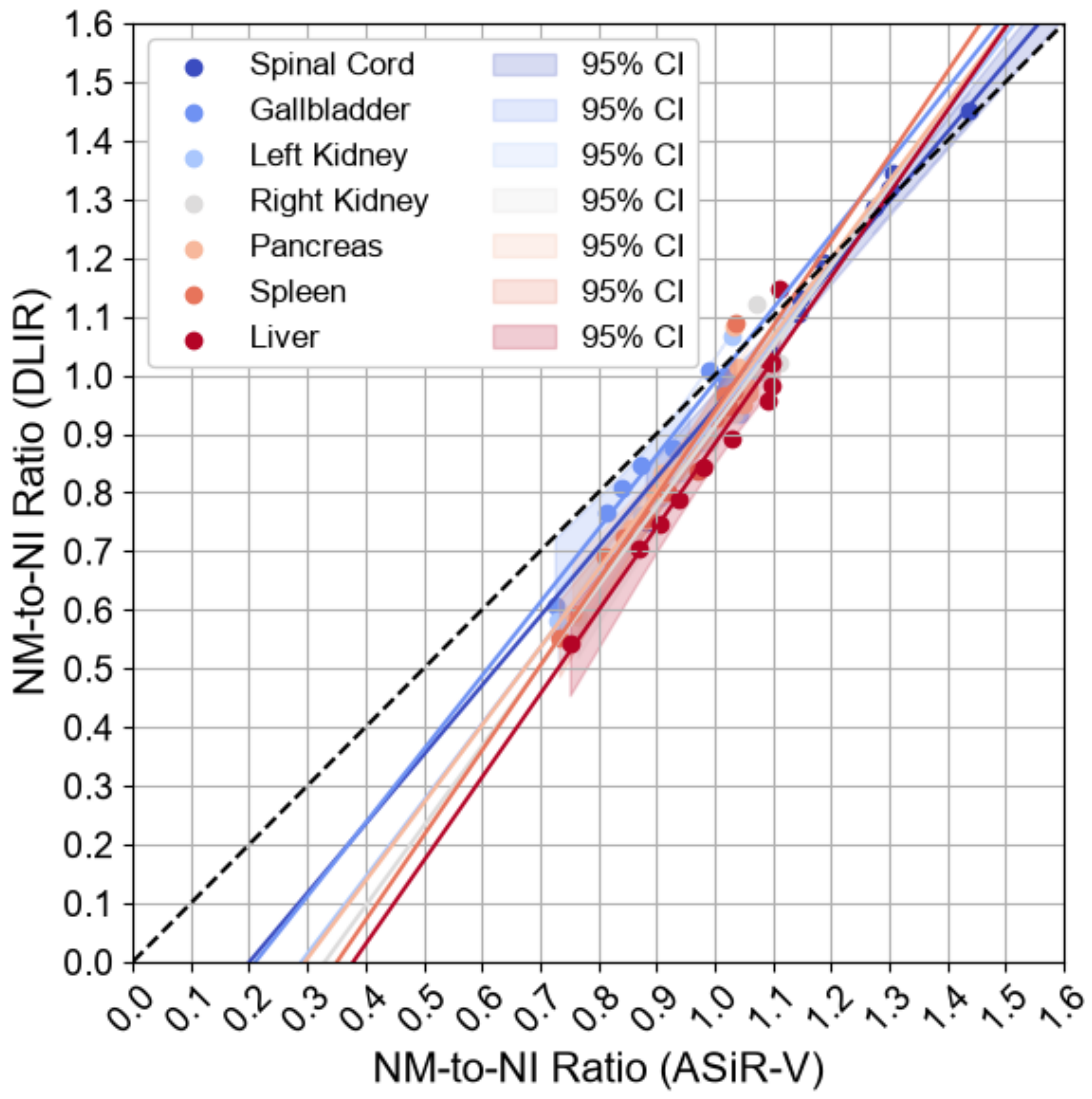

**Figure 10.** The relationship between the computed NM-to-NI ratios for the DLIR and ASiR-V images across all anatomical regions (scatter points) and the resulting linear lines-of-best-fit which can be used as a lookup table (dashed lines) and 95% CI (shaded region).

# 4. Discussion

## 4.1. Summary

This study was novel because it: 1) compared the resulting image noise across ASiR-V and DLIR reconstructions across varying NI values, 2) determined the approximate level of noise reduction at equivalent NI values, and 3) described the potential dose reductions of these findings at an organ-specific level. Several important findings should be noted.

The first is that sharper edges were seen in the noise image at the lowest NI setting in both the ASiR-V and DLIR images. One possible explanation for this is that there were either small acquisition (i.e., helical sampling or focal spot position) or reconstruction differences between the two repeat images which were not seen at the higher prescribed NI values due to the increased presence of distracting noise. Additional robustness analysis was done by performing two iterations of a segmentation erosion for each of the organ contours and re-computed the resulting fits. There were no significant differences seen in the resulting fitted values since the resulting values were within the 95% confidence intervals without erosion.

The second is that the $CTDI_{vol}$ related to NI by a power law with an exponent of -2.23 instead of -2, which is expected when considering the radiation dose and image noise relationship since noise is proportional to the inverse of the square root of dose so dose should be proportional to the inverse of noise squared[17]. One relevant observation here is that the ratio of the maximum to minimum mA falls monotonically from 1.53 at a prescribed NI of 5 to 1.18 at a prescribed NI of 30. This implies that the AEC is therefore performing less tube current modulation as the prescribed NI rises, which is a possible source of the organ-dependent behavior seen at high prescribed NI values.

The third is that, for both the ASiR-V and DLIR reconstructions, the exponent for the power law fit was less than 1 which indicates that the relationship between the computed NM and the prescribed NI is not linear. This relationship was descriptively less linear for the DLIR reconstruction due to its reduced exponents compared to the ASiR-V reconstruction for all tested organs (median of 0.66 compared to 0.80, respectively, with each organ showing the same trend). The slope discrepancy between the computed noise metric and the prescribed NI across both the ASiR-V and DLIR reconstructions comes from DLIR reconstruction being able to provide additional noise suppression at the equivalent NI compared to ASiR-V. Since the slope was found to be lower, its ability to suppress noise in inherently noisier images is stronger. Since the NI is defined as being approximately equal to the standard deviation in the central region of the image in a uniform phantom using a standard reconstruction algorithm according to the GE Healthcare technical reference manual, any deviation in this can result in non-linear results between the computed standard deviation between repeat images and the prescribed NI. For example, heterogeneous, hypointense, or hyperintense anatomy causes streaking artifacts which may contribute to noise inside a relatively homogeneous structure away from the region of concern. An interesting trend is that for the correlation between the computed NM and the prescribed NI is that, for the spinal cord, the spatial distribution of the scatter points is the same shape between the ASiR-V and DLIR reconstructions. This implies that there is a fundamental difference in the image acquisition at each prescribed NI value that DLIR cannot overcome. This observation could be due to the influence of the boney regions that surround the spinal cord which may lead to photon starvation and subsequently increased image noise.

The fourth is that when finding the equivalent prescribed NI that would result in the same computed NM when using the DLIR images instead of the ASiR-V images, a dependence on the baseline NI from the ASiR-V image was found with the resulting expected dose reduction for DLIR which was further organ-specific. The exact prescribed NI where this is a positive reduction is organ-dependent likely due to proximity to the surface of the patient, surrounding structures causing imaging artifacts, and overall size of the organ. Although the absolute dose reduction was not monotonically increasing across increasing prescribed NI values, the relative dose reduction was due to the lower doses used at higher NI settings.

Finally, the NM-to-NI ratio was found to decrease as the prescribed NI increased for both the ASiR-V and DLIR images. When comparing the NM-to-NI ratios between the DLIR and ASiR-V images, the NM-to-NI ratio for the DLIR images was found to be lower than for ASiR-V until an ASiR-V NM-to-NI ratio of 1.0 when the relationship switched. The exact point of switching was found to be approximately 1.0 to 1.3 (median ~1.2) depending on the organ. Lower NM-to-NI ratios imply that there is less image noise than what is being prescribed through the NI settings. An interesting trend can be seen for the ASiR-V plot where the NM-to-NI ratio is generally stable for prescribed NI values until a value of 10 before it starts to decrease. This is not seen in the DLIR images which demonstrates a monotonic decrease in the NM-to-NI ratio with increasing prescribed NI values. Since the NI is defined for "standard reconstructions", it can be inferred that the NM-to-NI ratio should stay constant when using a FBP reconstruction, though this should be tested in a future study. From this lens, the swap between the NM-to-NI ratio for DLIR being lower than ASiR-V to the reverse relationship could be due to DLIR being able to benefit unless the acquired image noise is too low, at which point the benefits of using DLIR, strictly from a noise reduction perspective, would be reduced. Added context is that one study found that DLIR reduces noise while maintaining texture and slightly enhancing spatial resolution[18] whereas another found that the noise power spectrum is higher for DLIR compared to ASiR-V at 50% strength[19].

Related prior work[20] to this study found that the noise as measured from the standard deviation in a sliding window increased in line with the spleen, kidneys, liver, and pancreas while this study found that the NM increased in line with the spleen, pancreas, kidneys, and liver for the ASiR-V reconstruction at an NI of 10. Several studies have investigated organ-specific dose estimations[21–23], but converting these studies to image noise implications remains underexplored[24]. Another study compared image noise across different organs between acquisitions organ dose modulation (ODM) and SmartmA using the standard deviation in a 15 x 15 pixel region of interest[25], however only the ratio of the noise between the two acquisition techniques was compared instead of the absolute values. A final set of studies developed a robust method for measuring image noise in CT scans[26,27], however they did not provide the same level of depth in their analysis as this study in terms of varying the prescribed noise index and resulting dose implications.

### 4.2. Limitations and Future Work

There were several limitations to this study. The first is that only the inferior portion of the chest, the entire abdomen, and the superior portion of the pelvis in the phantom were scanned during this study so future investigations for other anatomical regions should be conducted. The second is that only two different reconstruction methods were chosen since both ASiR-V and DLIR have the option to adjust their strengths which would allow for more identifiable trends to be seen. For example, when looking at the GE HealthCare operator manual for the Optima CT660, at the 50% ASiR-V strength used in this study, the NI factor to achieve the same image noise in a standard filtered back projection reconstruction is 1.70 and increases to 4.02 when using 100% ASiR-V strength, although this is just indicative since it is using a different platform. These values were determined using a homogenous water phantom but should be re-calculated on a per-organ basis using the scanner and phantom from this study. However, the reconstructions tested here were fair examples of clinically applicable reconstruction parameters. The third is that even though this phantom is more anatomically realistic than a traditional $CTDI_{vol}$ phantom, it still does not completely simulate a real human body, particularly of varying weights and heights which should be investigated further with novel phantoms. Another complication introduced in patients that is not represented in the phantom used in this study is physiological motion which would inherently increase the image noise due to shifting anatomy. This study only studied the implications of varying the prescribed NI for a single slice thickness so future studies should investigate this further. However, this can be difficult since the NI technically changes at varying slice thickness reconstructions even for the same acquisition since the total signal available is being reduced. For example, if an adjustment to the protocol was made to change the 5 mm slice thickness to 2.5 mm, then the actual NI would increase by $\sqrt{(5 / 2.5)} = \sqrt{2} = 1.41$, due to the reduced number of photons contributing to the signal within a given voxel. In line with this, different pitch factors should also be investigated since this study only evaluated the effects using a pitch of 1.375. Next, the scanner automatically switched to different focal spot sizes for the NI values of 5 and 7 which limits the direct scientific comparison between the different prescribed NI values. However, including these values increases the clinical relevance of the work since other clinics would see this switch in focal spot size as well. A brief robustness analysis was completed on the fitted power law curves for the NM to NI plots by removing the NI values of 5 and 7 and the fitted amplitude and exponents remained similar while staying within the 95% confidence intervals compared to using all the prescribed NI values. Finally, only two repeated acquisitions were obtained at each prescribed NI which allowed for a subtraction-based noise estimation, but this is not enough for a robust independent assessment of NM repeatability.

Future work should investigate how these results fit into the broader scope of the diagnostic imaging practice. For example, is a noise reduction of 10% noticeable when using the DLIR reconstruction compared to the ASiR-V reconstruction? What is the smallest change in the prescribed NI that is perceptible and can this be used to minimize patient dose on a per-organ study basis? Additionally, more in-depth analysis into the noise profiles of each organ using metrics like the noise power spectrum[28] (NPS), Monte Carlo[29], or radiomics[30,31] should be used in future studies to better understand how textural features change across prescribed NI, which may then affect radiologist interpretation of the images. This is especially important to study for CT since image noise is often correlated which may violate noise metrics based on the standard deviation within a region of interest[32].

# 5. Conclusion

This study explored the effects of varying acquired CT dose levels, in terms of the NI, on the reconstructed image noise, in terms of the NM, in an anthropomorphic phantom with realistic organ volumes and respective CT numbers using the 50% ASiR-V and DLIR reconstruction algorithms. Several observations were observed: 1) sharper edges were seen in the noise image at the lowest NI setting in both the ASiR-V and DLIR images, 2) $CTDI_{vol}$ related to NI by a power law with an exponent of -2.23, 3) the ratio of the maximum to minimum mA falls monotonically from 1.53 at a prescribed NI of 5 to 1.18 at a prescribed NI of 30, 4) the relationship between the computed NM and the prescribed NI is not linear, 5) the DLIR reconstruction provided additional noise suppression at the equivalent NI compared to ASiR-V, 6) dose reduction when using DLIR compared to ASiR-V is organ-specific, and 7) the NM-to-NI ratio was found to decrease as the prescribed NI increased for both the ASiR-V and DLIR images. Future studies should focus on expanding the reconstruction algorithm parameters investigated, studying other regions of the body using a similar methodology in an anthropomorphic phantom, and across different vendors and scanner models.

## References

1. Kopka L, Funke M, Breiter N, Hermann KP, Vosshenrich R, Grabbe E. [An anatomically adapted variation of the tube current in CT. Studies on radiation dosage reduction and image quality]. *Rofo*. 1995;163(5):383-387. doi:10.1055/s-2007-1016013

2. Lauritsch G, Haerer WH. Theoretical framework for filtered back projection in tomosynthesis. In: Hanson KM, ed. 1998:1127-1137. doi:10.1117/12.310839

3. Stiller W. Basics of iterative reconstruction methods in computed tomography: A vendor-independent overview. *European Journal of Radiology*. 2018;109:147-154. doi:10.1016/j.ejrad.2018.10.025

4. Prakash P, Kalra MK, Kambadakone AK, et al. Reducing abdominal CT radiation dose with adaptive statistical iterative reconstruction technique. *Invest Radiol*. 2010;45(4):202-210. doi:10.1097/RLI.ob013e3181dzfeec

5. Thibault J, Sauer KD, Bouman CA, Hsieh J. A three-dimensional statistical approach to improved image quality for multislice helical CT. *Medical Physics*. 2007;34(11):4526-4544. doi:10.1118/1.2789499

6. Kwon H, Cho J, Oh J, et al. The adaptive statistical iterative reconstruction-V technique for radiation dose reduction in abdominal CT: comparison with the adaptive statistical iterative reconstruction technique. *BJR*. 2015;88(1054):20150463. doi:10.1259/bjr.20150463

7. De Marco P, Origgi D. New adaptive statistical iterative reconstruction ASiR-V: Assessment of noise performance in comparison to ASiR. *J Applied Clin Med Phys*. 2018;19(2):275-286. doi:10.1002/acm2.12253

8. Koetzier LR, Mastrodicasa D, Szczykutowicz TP, et al. Deep Learning Image Reconstruction for CT: Technical Principles and Clinical Prospects. *Radiology*. 2023;306(3):e221257. doi:10.1148/radiol.221257

9. Dixon R, Anderson J, Bakalyar D, et al. *Comprehensive Methodology for the Evaluation of Radiation Dose in X-Ray Computed Tomography*. AAPM; 2010. doi:10.37206/109

10. Ono K, Asada Y. Analysis of material composition and attenuation characteristics of anthropomorphic torso phantoms for dosimetry using dual energy CT technology. *Phys Eng Sci Med*. 2025;48(2):675-683. doi:10.1007/s13246-025-01533-1

11. Zarei M, Abadi E, Segars WP, Samei E. Coronary stenosis quantification in cardiac computed tomography angiography: multi-factorial optimization of image quality and radiation dose. *JMI*. 2023;10(6):063502. doi:10.1117/1.JMI.10.6.063502

12. Wasserthal J, Breit HC, Meyer MT, et al. TotalSegmentator: Robust Segmentation of 104 Anatomic Structures in CT Images. *Radiology: Artificial Intelligence*. 2023;5(5):e230024. doi:10.1148/ryai.230024

13. Isensee F, Jaeger PF, Kohl SAA, Petersen J, Maier-Hein KH. nnU-Net: a self-configuring method for deep learning-based biomedical image segmentation. *Nat Methods*. 2021;18(2):203-211. doi:10.1038/s41592-020-01008-z

14. Fedorov A, Beichel R, Kalpathy-Cramer J, et al. 3D Slicer as an image computing platform for the Quantitative Imaging Network. *Magnetic Resonance Imaging*. 2012;30(9):1323-1341. doi:10.1016/j.mri.2012.05.001

15. Tian X, Samei E. Accurate assessment and prediction of noise in clinical CT images. *Medical Physics*. 2016;43(1):475-482. doi:10.1118/1.4938588

16. Steel RGD, Torrie JH. *Principles and Procedures of Statistics: With Special Reference to the Biological Sciences*. McGraw-Hill; 1960.

17. Scott AW, Alancherry ASP, Lee C, Eastman E, Zhou Y. Computed tomography dose index determination in dose modulation prospectively involving the third-generation iterative reconstruction and noise index. *J Applied Clin Med Phys*. 2024;25(4):e14167. doi:10.1002/acm2.14167

18. Greffier J, Si-Mohamed S, Frandon J, et al. Impact of an artificial intelligence deep-learning reconstruction algorithm for CT on image quality and potential dose reduction: A phantom study. *Med Phys*. 2022;49(8):5052-5063. doi:10.1002/mp.15807

19. Yoo YJ, Choi IY, Yeom SK, et al. Evaluation of Abdominal CT Obtained Using a Deep Learning-Based Image Reconstruction Engine Compared with CT Using Adaptive Statistical Iterative Reconstruction. *J Belg Soc Radiol*. 2022;106(1):15. doi:10.5334/jbsr.2638

20. Wang S, Medrano MJ, Imran AAZ, et al. Automated estimation of individualized organ-specific dose and noise from clinical CT scans. *Phys Med Biol*. 2025;70(3):035014. doi:10.1088/1361-6560/ada67f

21. Tzanis E, Damilakis J. A machine learning-based pipeline for multi-organ/tissue patient-specific radiation dosimetry in CT. *Eur Radiol*. 2024;35(2):919-928. doi:10.1007/s00330-024-11002-0

22. Myronakis M, Stratakis J, Damilakis J. Rapid estimation of patient-specific organ doses using a deep learning network. *Medical Physics*. 2023;50(11):7236-7244. doi:10.1002/mp.16356

23. Schmidt TG, Wang AS, Coradi T, Haas B, Star-Lack J. Accuracy of patient-specific organ dose estimates obtained using an automated image segmentation algorithm. *J Med Imag*. 2016;3(4):043502. doi:10.1117/1.JMI.3.4.043502

24. Pace E, Caruana CJ, Bosmans H, Cortis K, D'Anastasi M, Valentino G. An inventory of patient-image based risk/dose, image quality and body habitus/size metrics for adult abdomino-pelvic CT protocol optimisation. *Physica Medica*. 2024;125:103434. doi:10.1016/j.ejmp.2024.103434

25. Gandhi D, Crotty DJ, Stevens GM, Schmidt TG. Technical Note: Phantom study to evaluate the dose and image quality effects of a computed tomography organ-based tube current modulation technique. *Medical Physics*. 2015;42(11):6572-6578. doi:10.1118/1.4933197

26. Christianson O, Winslow J, Frush DP, Samei E. Automated Technique to Measure Noise in Clinical CT Examinations. *American Journal of Roentgenology*. 2015;205(1):W93-W99. doi:10.2214/AJR.14.13613

27. Ahmad M, Jacobsen MC, Thomas MA, Chen HS, Layman RR, Jones AK. A Benchmark for automatic noise measurement in clinical computed tomography. *Medical Physics*. 2021;48(2):640-647. doi:10.1002/mp.14635

28. S J Riederer, N J Pelc, D A Chesler. The noise power spectrum in computed X-ray tomography. *pmb*. 1978;23(3):446-454. doi:10.1088/0031-9155/23/3/008

29. Lee C, Liu J, Griffin K, Folio L, Summers RM. Adult patient-specific CT organ dose estimations using automated segmentations and Monte Carlo simulations. *Biomed Phys Eng Express*. 2020;6(4):045016. doi:10.1088/2057-1976/ab98e6

30. Lambin P, Rios-Velazquez E, Leijenaar R, et al. Radiomics: Extracting more information from medical images using advanced feature analysis. *European Journal of Cancer*. 2012;48(4):441-446. doi:10.1016/j.ejca.2011.11.036

31. Mackin D, Fave X, Zhang L, et al. Measuring Computed Tomography Scanner Variability of Radiomics Features: *Investigative Radiology*. 2015;50(11):757-765. doi:10.1097/RLI.0000000000000180

32. Hsieh SS, Leng S, Yu L, Huber NR, McCollough CH. A minimum SNR criterion for computed tomography object detection in the projection domain. *Med Phys*. 2022;49(8):4988-4998. doi:10.1002/mp.15832

# Supplementary Materials

*Derivation 1:*

$$\mathrm{NM_{ASiR\text{-}V}} = a_1(\mathrm{NI_{ASiR\text{-}V}})^{b_1}$$

$$\mathrm{NM_{DLIR}} = \mathrm{NM_{ASiR\text{-}V}}$$

$$\rightarrow a_{\mathrm{NM}}(\mathrm{NM_{ASiR-V}})^{b_{\mathrm{NM}}} = a_1(\mathrm{NI_{ASiR\text{-}V}})^{b_1}$$

$$\rightarrow a_{\mathrm{NM}}\left[a_1(\mathrm{NI_{DLIR}})^{b_1}\right]^{b_{\mathrm{NM}}} = a_1(\mathrm{NI_{ASiR\text{-}V}})^{b_1}$$

$$\therefore \mathrm{NI_{DLIR}} = \left(\frac{{a_1}^{1-b_{\mathrm{NM}}}}{a_{\mathrm{NM}}}\right)^{1/b_1 b_{\mathrm{NM}}} (\mathrm{NI_{ASiR-V}})^{1/b_{\mathrm{NM}}}$$

*Derivation 2:*

$$\mathrm{CTDI_{vol}} = a_{\mathrm{CTDI_{vol}}}(\mathrm{NI})^{b_{\mathrm{CTDI_{vol}}}}$$

$$\rightarrow \mathrm{CTDI_{vol,ASiR-V}} = a_{\mathrm{CTDI_{vol}}}(\mathrm{NI_{ASiR-V}})^{b_{\mathrm{CTDI_{vol}}}}$$

$$\therefore \mathrm{CTDI_{vol,DLIR}} = a_{\mathrm{CTDI_{vol}}}(\mathrm{NI_{DLIR}})^{b_{\mathrm{CTDI_{vol}}}} = a_{\mathrm{CTDI_{vol}}}\left[\left(\frac{{a_1}^{1-b_{\mathrm{NM}}}}{a_{\mathrm{NM}}}\right)^{1/b_1 b_{\mathrm{NM}}} (\mathrm{NI_{ASiR-V}})^{1/b_{\mathrm{NM}}}\right]^{b_{\mathrm{CTDI_{vol}}}}$$

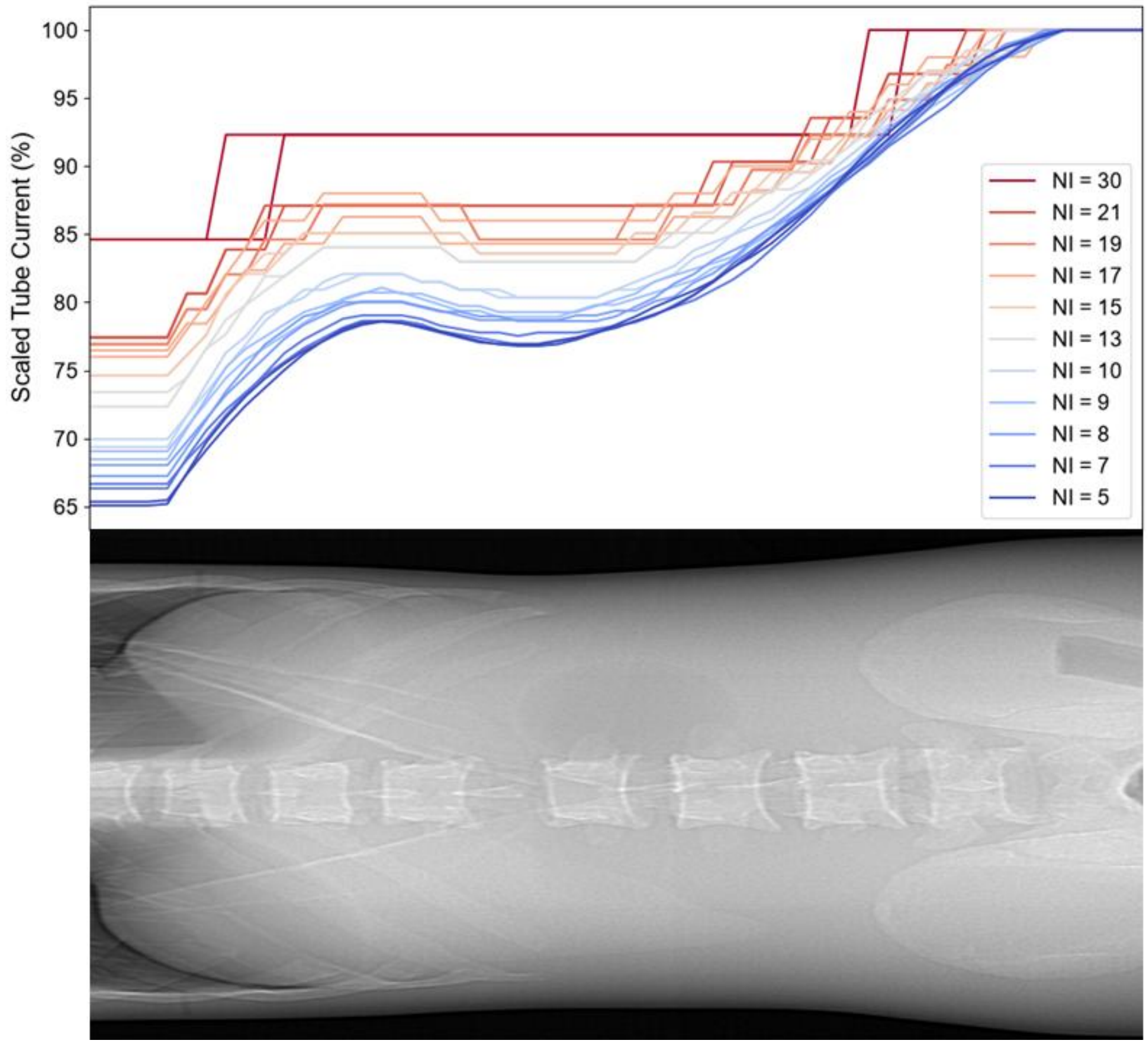

**Figure S1.** The effects of varying NI on the automatic exposure control through varying tube current across the length of the phantom scaled by the maximum tube current. Loss in the number of available tube current settings and their dynamic range as NI increases may be an additional factor to consider when analyzing the results.

**Table S1.** Descriptive statistics of each organ in the phantom including its volume, approximate depth at the center of the organ, and the mean CT numbers for both the ASiR-V and DLIR reconstructions.

| | **Volume ($cm^3$)** | **Approximate Depth (cm)** | **Mean CT Number (HU, ASiR-V)** | **Mean CT Number (HU, DLIR)** |
|---|---|---|---|---|
| Spinal Cord | 43 | 4 | 8 | 8 |
| Gallbladder | 15 | 5 | 18 | 18 |
| Left Kidney | 88 | 6 | 26 | 26 |
| Right Kidney | 88 | 7 | 26 | 26 |
| Pancreas | 41 | 7 | 26 | 27 |
| Spleen | 141 | 4 | 49 | 50 |
| Liver | 1086 | 8 | 68 | 68 |

**Table S2.** Relationship between the computed NM and prescribed NI for each anatomical region for both the ASiR-V and DLIR reconstructions for the equation $a(NI)^b$ with the accompanying 95% confidence intervals (CI).

| | **ASiR-V** | | **DLIR** | |
|---|---|---|---|---|
| | a (95% CI) | b (95% CI) | a (95% CI) | b (95% CI) |
| Spinal Cord | 2.89 (2.18, 3.82) | 0.66 (0.55, 0.77) | 3.60 (2.68, 4.83) | 0.56 (0.45, 0.68) |
| Gallbladder | 1.43 (1.14, 1.77) | 0.82 (0.74, 0.91) | 1.77 (1.43, 2.20) | 0.72 (0.64, 0.81) |
| Left Kidney | 1.56 (1.31, 1.86) | 0.80 (0.73, 0.87) | 1.89 (1.60, 2.24) | 0.68 (0.62, 0.75) |
| Right Kidney | 1.64 (1.37, 1.98) | 0.80 (0.73, 0.87) | 2.09 (1.69, 2.58) | 0.66 (0.58, 0.75) |
| Pancreas | 1.52 (1.27, 1.81) | 0.82 (0.75, 0.89) | 1.85 (1.52, 2.27) | 0.70 (0.62, 0.78) |
| Spleen | 1.54 (1.33, 1.79) | 0.80 (0.74, 0.86) | 2.02 (1.74, 2.34) | 0.65 (0.59, 0.70) |
| Liver | 1.71 (1.44, 2.03) | 0.78 (0.72, 0.85) | 2.27 (1.88, 2.74) | 0.61 (0.54, 0.69) |

**Table S3.** Relationship between the computed NM for the DLIR reconstruction compared to the ASiR-V reconstruction at equivalent prescribed NI values for the equation $a(NM_{ASiR\text{-}V})^b$ with the accompanying 95% confidence intervals (CI).

| | **a (95% CI)** | **b (95% CI)** |
|---|---|---|
| Spinal Cord | 1.43 (1.22, 1.67) | 0.86 (0.80, 0.92) |
| Gallbladder | 1.30 (1.13, 1.50) | 0.87 (0.82, 0.93) |
| Left Kidney | 1.29 (1.19, 1.41) | 0.85 (0.82, 0.89) |
| Right Kidney | 1.37 (1.22, 1.55) | 0.83 (0.79, 0.88) |
| Pancreas | 1.29 (1.13, 1.48) | 0.86 (0.81, 0.91) |
| Spleen | 1.42 (1.29, 1.56) | 0.81 (0.77, 0.85) |
| Liver | 1.48 (1.32, 1.65) | 0.79 (0.74, 0.83) |

**Table S4.** Relationship between the DLIR NM-to-NI ratio and the ASiR-V NM-to-NI ratio for the equation a(NM-to-NI $ratio_{ASiR-V}$) + b with the accompanying 95% confidence intervals (CI).

| | a (95% CI) | b (95% CI) | $r^2$ |
|---|---|---|---|
| Spinal Cord | 1.18 (1.10, 1.26) | -0.24 (-0.34, -0.13) | 0.99 |
| Gallbladder | 1.25 (0.89, 1.61) | -0.26 (-0.60, 0.07) | 0.87 |
| Left Kidney | 1.31 (1.04, 1.57) | -0.38 (-0.63, -0.13) | 0.93 |
| Right Kidney | 1.36 (1.09, 1.64) | -0.45 (-0.73, -0.17) | 0.93 |
| Pancreas | 1.33 (1.01, 1.64) | -0.39 (-0.70, -0.09) | 0.91 |
| Spleen | 1.45 (1.16, 1.74) | -0.51 (-0.78, -0.23) | 0.93 |
| Liver | 1.42 (1.15, 1.70) | -0.54 (-0.81, -0.26) | 0.94 |